\documentclass[12pt]{iopart}

\usepackage{iopams}
\usepackage{graphicx}
\begin{document}

\title[Comparison between two mobile absolute gravimeters:\\optical versus atomic interferometers]{Comparison between two mobile absolute gravimeters: optical versus atomic interferometers}

\author{S.~Merlet, Q. Bodart, N.~Malossi\footnote{Present address: Dipartimento di Fisica Enrico Fermi, Universita di Pisa, L. Pontecorvo 3, I-56127, ITALY}, A.~Landragin and F.~Pereira Dos Santos}

\address{LNE-SYRTE, Observatoire de Paris, CNRS et UPMC, 61 avenue de l'Observatoire, 75014 Paris, FRANCE}
\ead{franck.pereira@obspm.fr}

\author{O.~Gitlein, L.~Timmen}
\address{Institut f\"{u}r Erdmessung, Leibniz Universit\"{a}t of Hannover, Scheinderberg 50, 30167 Hannover, GERMANY}

\begin{abstract}
We report a comparison between two absolute gravimeters~: the LNE-SYRTE cold atoms gravimeter and FG5\#220 of Leibniz Universit\"{a}t of Hannover. They rely on different principles of operation~: atomic and optical interferometry. Both are movable which enabled them to participated to the last International Comparison of Absolute Gravimeters (ICAG'09) at BIPM. Immediately after, their bilateral comparison took place in the LNE watt balance laboratory and showed an agreement of $(4.3\pm6.4)~\mu$Gal.

\end{abstract}

\maketitle

\section{Introduction}
Over the last two decades inertial sensors based on atom interferometry have been realized. In particular, as described in \cite{Kasevich}, cold atoms gravimeters can reached performances comparable to "classical" corner cube gravimeters both in terms of sensitivity \cite{Legouet, Muller} and accuracy. The first and unique comparison between atomic and optical gravimeters \cite{Peters} has shown an agreement between the sensors ($(7\pm7)~\mu$Gal\footnote{$1~\mu$Gal=$10^{-8}$m.s$^{-2}$} difference). In this paper, we present the result of a comparison, realized between the cold atom gravimeter (CAG) developed by LNE-SYRTE in the frame of the French watt balance project \cite{Geneves} and the FG5\#220 of Leibniz Universit\"{a}t of Hannover (LUH) \cite{Timmen}. Both rely on the measurement of the trajectory of free falling bodies (corner cube for FG5 and $^{87}$Rb atoms for CAG). Unlike the situation described in \cite{Peters}, both sensors are mobile which makes regular comparisons at various sites possible. Such comparisons between instruments based on different technologies are of fundamental interest for accurate metrology of $g$. This motivated the participation of both devices to ICAG'09 and the subsequent bilateral comparison presented in this paper. For this purpose, both sensors were moved from BIPM to the gravimetry room (\emph{GR}) of the LNE watt balance laboratory \cite{Merlet}, where they performed simultaneous gravity measurements.

\section{Experimental setups}

\begin{figure}[h!]
    \centering
    \includegraphics[width=7 cm, angle=0]{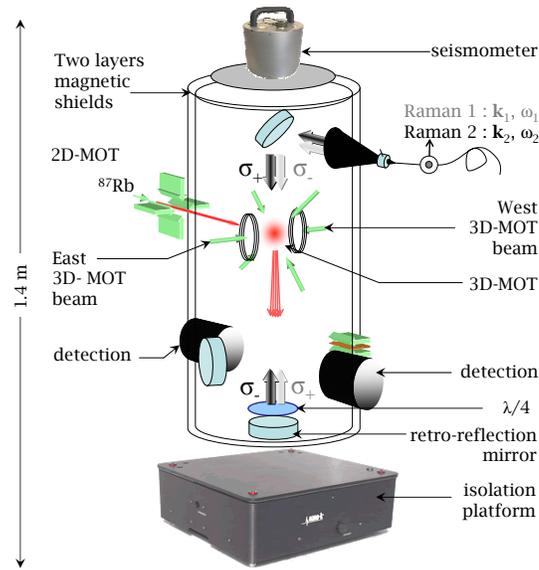}
\caption{\small{Scheme of the CAG set-up. The drop chamber made of titanium is placed onto a passive isolation platform. Atoms are first trapped in a Magneto Optical Trap (MOT), cooled with optical molasses and released. During the free fall, the interferometer is realized with vertical Raman laser beams. The $g$ measurement is determined from the interferometer phase shift.}}
    \label{setup}
\end{figure}

The CAG is an improved version of a prototype described in \cite{Legouet} which reached a short term sensitivity to acceleration of $1.4\times10^{-8}g$ at 1~s. It is composed of three parts~: a dropping chamber on its isolation platform (fig. \ref{setup}), a compact optical bench (60$\times90$~cm$^{2}$) \cite{Cheinet} and two 2~m lab racks for the electronic control. The Earth's acceleration measurement is deduced from the phase difference between the two paths of an interferometer realized with cold atoms.
The FG5 absolute gravimeter of LUH is a state-of-the-art commercial gravimeter which is essentialy a modified Mach-Zender "in-line" interferometer as described in \cite{Niebauer}.

\section{Results}

\begin{figure}[h!]
    \centering
    \includegraphics[width=5 cm, angle=-90]{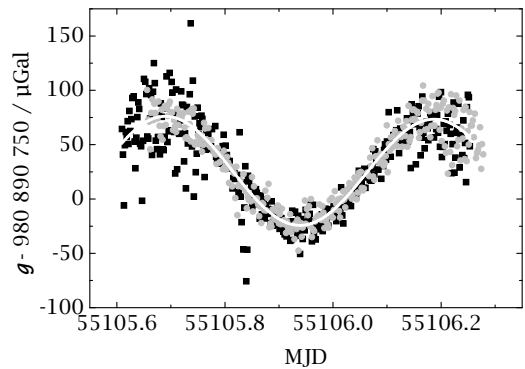}
\caption{\small{Earth's gravity variation $g$ during the night from the $1^{st}$ to the $2^{d}$ of october 2009 on site \emph{GR} at LNE. Dots represent average data over 2 min 30 s (black squares: FG5\#220, gray circles: CAG). Tidal variation is plotted as a white line on the data.}}
    \label{signaux}
\end{figure}

\begin{figure}[h!]
    \centering
    \includegraphics[width=5 cm, angle=-90]{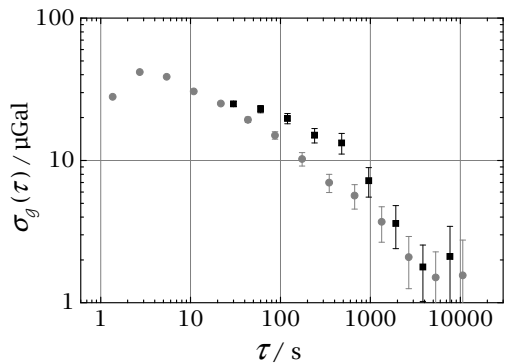}
\caption{\small{Allan standard deviation of the corrected signals: FG5\#220 (black squares), CAG (gray circles).}}
    \label{var}
\end{figure}

Gravimeters measured simultaneously all the night in the well characterized \emph{GR} room \cite{Merlet}. The FG5\#220 was located on point \emph{GR}$_{29}$ with one drop per 30~s. The result transferred at 120~cm is reported in Table \ref{tab1}. The CAG was on point \emph{GR}$_{40}$, measuring at the high cycling rate of 3~Hz. Its result, also transferred at 120~cm, is reported in Table \ref{tab1}. The two points \emph{GR}$_{40}$ and \emph{GR}$_{29}$ are 2.12~m apart and the tie between them, obtained with a Scintrex CG5, is \emph{g}$_{GR40}$-\emph{g}$_{GR29}$ = $(6.5\pm1.0)~\mu$Gal at the height of 120~cm \cite{Merlet}. Transferred on point \emph{GR}$_{40}$ at $120~cm$, the difference between the devices is $(4.3\pm6.4)~\mu$Gal. The $g$ measurements uncorrected from tides are displayed on figure \ref{signaux}. The stability is characterized by the Allan standard deviation of the tide-corrected $g$ measurements (fig. \ref{var}).

\begin{table}
\centering
\begin{tabular}{c c c c c }
\hline
device & point  &  $g$  & \emph{U}(k=1) & \emph{s}$_{gm}$  \\
 & & /$\mu$Gal& /$ \mu$Gal& /$\mu$Gal \\
   \hline
   \hline
CAG & \emph{GR}$_{40}$  & 980~890~744.8 & 5.9  & 0.7 \\
FG5\#220 & \emph{GR}$_{29}$  & 980~890~742.6 &  2.2 & 1.0 \\
CG5 & \emph{GR}$_{40}$-\emph{GR}$_{29}$  & ~~~~~~~~~~~~~~6.5 & 1.0 & 0.1 \\
$\mathbf{CAG-FG5\#220}$ & $\mathbf{\emph{GR}_{40}}$  & ~~~~~~~~~~~ $\mathbf{-4.3}$ & $\mathbf{6.4}$ & $\mathbf{1.6}$ \\
\hline
\end{tabular}
\caption{Gravity results at $120~cm$ height. }
\label{tab1}
\end{table}

Despite different vibration isolation systems and repetition rates, the signal dispersions are similar except during first hours of the comparison, as can be seen on the figure \ref{var}. Measurements are found to be less noisy after midnight due to the drastic reduction of the human activity in the surrounding industrial area. At best, the FG5 drop scatter is $16~\mu$Gal. The CAG's $g$ determination is based on four successive configurations measurement in order to reject the bias due to the two photon light shift \cite{Gauguet}. This degrades the sensitivity by a factor $\sqrt{10}$. Better sensitivity could also be obtained with the FG5 if performing with one drop per 10~s rather than 30~s chosen to preserve the device.

\vspace{1cm}

Rotating the CAG by $180^{\circ}$ around the vertical axis, we measured a Coriolis shift of $(6.5\pm0.5)~\mu$Gal. Varying the temperature of the atoms enabled us to evaluate the bias due to wavefront abberations \cite{Fils} to $(3.0\pm3.0)~\mu$Gal. During this comparison, the lack of rigidity of the mechanical structure resulted in a relatively large vertical alignment bias of $(4.5\pm4.5)~\mu$Gal. The final accuracy was $5.9~\mu$Gal.
In \cite{Niebauer} an error analysis of the FG5 system lead to a total uncertainty of $1.1~\mu$Gal. From numerous comparisons with other absolute gravimeters since 2002, the LUH group estimates the accuracy of their device to be $2.0~\mu$Gal \cite{Timmen, Timmen2}. The \emph{g} result measured by FG5\#220 on point \emph{GR}$_{29}$ agrees with the mean of previous measurements performed with other FG5s on the same point, in October 2006 \cite{Merletc} (difference of $(1.9\pm2.9)~\mu$Gal).

\section{Conclusion and discussion}

We have compared two different portable absolute gravimeters and found an agreement of $(4.3\pm6.4)~\mu$Gal. More such comparisons will be realized in the future while striving to improve the accuracy of the CAG down to 1~$\mu$Gal. Already, the vertical alignment bias has been reduce to $(0.0\pm0.5)~\mu$Gal.  Future comparisons will benefit from the mobility of atomic sensors as described here. Transportability is an important and original feature of the CAG, which is necessary for regular participation to comparison campaigns. Nevertheless, CAG is still a laboratory device but such comparisons as the "field" gravimetric measurements would benefit from development of a more compact gravimeter as described in \cite{Bodart}.

\ack

We would like to thank IFRAF and ESF (EuroQUASAR) for financial support. Q.B thanks CNES for supporting his work.

\end{document}